\documentstyle[11pt,aaspp4,flushrt]{article} \parskip 0pt

\def\[#1]{\cite{#1}}
\def\eq#1{\begin{equation} #1 \end{equation}}
\def\eqarray#1{\begin{eqnarray} #1 \end{eqnarray}}
\def\non{\nonumber \\}

\def\DnuD     {\hbox{$\Delta\nu_D$}}
\def\DnuB     {\hbox{$\Delta\nu_B$}}
\def\Dnum     {\hbox{$\Delta\nu_m$}}
\def\D        {\hbox{${\cal D}$}}
\def\R        {\hbox{${\cal R}$}}
\def\RI       {\hbox{$R_1$}}
\def\Rc       {\hbox{$R_c$}}
\def\xb       {\hbox{$x_B$}}
\def\k        {\hbox{$\kappa$}}
\def\kM       {\hbox{$\kappa^{\Delta m}$}}
\def\km       {\hbox{$\kappa_m$}}
\def\kc       {\hbox{$\kappa_c$}}
\def\kl       {\hbox{$\kappa_l$}}
\def\ko       {\hbox{$\kappa_0$}}
\def\dko      {\hbox{$\kappa'_0$}}
\def\koo      {\hbox{$\kappa_0^0$}}
\def\npm      {\hbox{$\nu_\pm$}}
\def\vpm      {\hbox{$v^\pm$}}
\def\Onu      {\hbox{$\Omega(\nu)$}}
\def\about    {\hbox{$\sim$}}
\def\bfe      {\hbox{\boldmath $\epsilon$}}
\def\bfS      {\hbox{\boldmath $S$}}
\def\bfK      {\hbox{\boldmath $K$}}
\def\dot      {\hbox{\boldmath $\cdot$}}
\def\PI       {\hbox{$\bf \Pi$}}
\def\half     {\hbox{$1\over2$}}
\def\fourth   {\hbox{$1\over4$}}
\def\tm       {\hbox{$\tau_m$}}
\def\tp       {\hbox{$\tau_\pi$}}
\def\Ith      {\hbox{$I_{\rm th}$}}

\begin{document}

\rightline{To appear in ApJ {\bf 504}, September 1, 1998}
\title
            {POLARIZATION OF ASTRONOMICAL MASER RADIATION. IV.\\
                       CIRCULAR POLARIZATION PROFILES}

\author
                               {Moshe Elitzur\\
                   Department of Physics and Astronomy, \\
             University of Kentucky, Lexington, KY 40506-0055 \\
                              moshe@pa.uky.edu}

\begin                           {abstract}

Profile comparison of the Stokes parameters $V$ and $I$ is a powerful tool for
maser data analysis, providing the first direct methods for unambiguous
determination of (1) the maser saturation stage, (2) the amplification optical
depth and intrinsic Doppler width of unsaturated masers, and (3) the comparative
magnitudes of Zeeman splitting and Doppler linewidth.  Circular polarization
recently detected in OH 1720 MHz emission from the Galactic center appears to
provide the first direct evidence for maser saturation.
\end{abstract}


\section{INTRODUCTION}

Few properties of astronomical masers are determined directly by observations,
most are inferred indirectly.  Foremost among the latter is the maser saturation
stage. Saturation has a significant impact on maser growth, so determining
whether a maser is saturated ($J > J_s$, where $J$ is the angle-averaged
intensity and $J_s$ is the saturation intensity) or not is usually a
precondition for analysis of the observations.  Unfortunately, this crucial
issue is not convincingly settled.  Strong masers are generally believed to be
saturated, but in most cases the evidence is less than compelling as it relies
primarily on plausibility arguments rather than quantitative tests (see e.g.\
\[Book], \S 8.6).  This unsatisfactory situation reflects a fundamental
difficulty --- neither $J$ nor $J_s$ is directly measurable.  The saturation
parameter $J_s$ is a theoretical quantity, determined only within the context of
a given pumping scheme.  And because maser radiation is highly beamed, $J =
I\Omega/4\pi$ so this quantity, too, cannot be directly measured; the intensity
$I$ is measurable when the maser is resolved, but the beaming angle $\Omega$ is
unobservable.  Similarly, the amplification optical depth has never been
directly determined for any maser that amplifies its own radiation.

Recent VLA observations of OH 1720 MHz masers near the Galactic center by
\[YZ96] open up new possibilities for direct determination of some maser
properties. Significant circular polarization (upward of 20\%) is detected in
various spectral features, and the right- and left-hand components coincide on
the sky, as expected from the Zeeman effect. Furthermore, the spectral shape of
the Stokes parameter $V$ follows an antisymmetric S-curve with sharp reversal at
line center, the typical profile for Zeeman shift \DnuB\ much smaller than the
Doppler linewidth \DnuD.  Similar results were previously reported for H$_2$O
masers in star-forming regions by \[FbG] and for OH 1612 MHz masers in OH/IR
stars by \[ZFix], but the polarization was lower and the quality of the data not
nearly as high.

The general maser polarization solution was recently derived for arbitrary
values of
\eq{
                         \xb = {\DnuB \over \DnuD}
}
(Elitzur 1996, hereafter \[E96]) and the solution properties at $\xb \ll 1$
closely match the observed circular polarization. Here I show that a comparative
analysis of the spectral profiles of $I$ and $V$, two measurable independent
maser intensities, offers direct determination of various maser properties, in
particular the saturation stage.  The analysis is readily performed with the aid
of the ratio profile
\eq{\label{R0}
              \R(\nu) = {V(\nu) \over I'(\nu)} = {v(\nu)\over I'(\nu)/I(\nu)}
}
where the prime denotes derivative with respect to frequency and $v = V/I$ is
the fractional circular polarization.  When $\xb \ll 1$, spectral analysis of
\R\ offers intrinsic sensitivity of order \xb\ and has long been an important
tool in studies of the Zeeman effect of thermal radiation (see e.g.\ \[Trol]).
In that case \R\ is constant across the spectral line and its magnitude
determines the magnetic field along the line of sight.  This constancy of \R\
follows from some simple, general symmetry arguments as shown by \[Cru] (see
also \S3 below). However, maser exponential amplification during unsaturated
growth destroys both the underlying symmetry and the constancy of \R, the
saturation process restores both. The key to the different behavior, and
\R-profiles, in the two regimes is the narrowing of the maser line during
unsaturated amplification and its rebroadening during saturation.

The important differences between thermal and maser polarization are discussed
in detail below.  For completeness, some basic elements of the polarization
theory developed in \[E96] are reproduced in \S2.  The \R\ profile is discussed
in \S3 for thermal radiation and in \S4 for maser radiation when $\xb \ll 1$.
In \S5, circular polarization for fully resolved Zeeman patterns, $\xb > 1$, is
discussed.  The implications for observations are discussed in detail in \S6.

\section{GENERALITIES}

A full description of electromagnetic radiation involves the 4-vector of its
Stokes parameters $\bfS = (I, Q, U, V)$.  The general transfer equation for
\bfS\ is
\eq{\label{basic}
               {d\bfS\over dl} = \bfe + \bfK\dot\bfS,
}
where \bfe\ is the 4-vector analog of the familiar volume emission coefficient
and \bfK\ is the matrix analog of the absorption coefficient. In the case of
line radiation, \bfK\ was derived in \[Lit75] and is reproduced here using the
notation of \[E96].  For any spin transition, the radiative interactions with
photons of polarization $\Delta m$ $(= +1, 0, -1)$ are characterized by an
absorption coefficient \kM.  Introduce
\eq{
                           \k^1 = \half(\k^+ + \k^-),
}
the mean absorption coefficient for $\Delta m = \pm1$ transitions, and the three
linear combinations
\eqarray{\label{kappas}
    \km &= &\half[\k^1(1 + \cos^2\theta) + \k^0\sin^2\theta]  \non
    \kl &= &\half(\k^1 - \k^0)\sin^2\theta                    \non
    \kc &= &\half(\k^+ - \k^-)\cos\theta,
}
where $\theta$ is the angle of the radiation propagation direction from the
quantization axis, taken here along the magnetic field.  Then the transfer
matrix for line radiation is
\eq{\label{R}
     \bfK = \pmatrix     {\km & \kl &  0  & \kc         \cr
                          \kl & \km &  0  &  0          \cr
                           0  &  0  & \km &  0          \cr
                          \kc &  0  &  0  & \km         \cr}
}

The polarization of the 4-vector \bfS\ is characterized by $\PI = (q, u, v)$,
the 3-vector of its normalized Stokes parameters $q = Q/I$, $u = U/I$, $v =
V/I$. It is easy to verify that the transfer matrix \bfK\ preserves full
polarization,  $|\PI| = 1$. From Maxwell's equations, radiation is always
emitted fully polarized and for any intensity $I$ there is an ensemble of
configurations of \bfS\ (polarization modes) that differ from each other only by
the directions of their corresponding unit vectors \PI, i.e., the sense of
polarization.  For each mode, \PI\ remains a unit vector and radiative transfer
only rotates it, transferring polarization between the linear and circular
polarizations of individual modes. The polarization vector of the entire
radiation field at intensity $I$ is obtained from the ensemble average of all
polarization modes.  This average differs from zero, and the observed radiation
is polarized, only when certain modes are favored.

The emission term for the mode with initial polarization $\PI_0$ can be written
as
\eq{
       \bfe = \km\bfS_0\,, \qquad \hbox{where \ } \bfS_0 = S_0(1, \PI_0)
}
and $S_0$ is the standard source function. The ensemble average of these source
terms has the common intensity $S_0$, and when the individual polarizations
point at random directions in $q,u,v$ space it also has zero polarization.  This
is usually the case for radiation generated in spontaneous decays. The radiative
transfer equations admit the formal solution for the intensity of individual
modes
\eq{\label{formal}
                    I = S_0 e^{\tm + \tp},
}
where
\eq{\label{taus}
                  \tm = \int \km dl, \qquad
                  \tp = \int (\kl q + \kc v) dl.
}
The optical depth \tp\ describes the rotation of the mode polarization while the
growth of its intensity is dominated by \tm, which is always larger than \tp.
Modes that have \tp\ = 0 do not rotate, providing a stationary polarization
configuration.

\section{THERMAL CIRCULAR POLARIZATION}

In the case of thermal emission, the emerging radiation samples no more than
\about\ one optical depth into the source and its properties primarily reflect
those of the source terms. Therefore, only the polarization of the emission
terms need be considered; the subsequent effect of radiative transfer is
generally only a perturbation.

In the presence of a magnetic field, thermal emission is circularly polarized
and the polarization can be derived from rather general considerations (e.g.,
\[Cru]).  The energies of magnetic sub-levels are shifted by the Zeeman effect,
splitting each line into components with $\Delta m = -1, 0, +1$ (where $\Delta
m$ refers to the $m$-change in absorption for compatibility with the previous
section). The standard Zeeman emission pattern is comprised of three components
centered on the frequencies $\nu_0 - \Delta m\DnuB$, each one with appropriate
polarization properties. From symmetry, for $m$-independent line excitations the
intensities of the three components are related via
\eq{\label{shift}
    I^{\pm}(x) = I^0(x \pm \xb) \simeq I^0(x) \pm \xb{dI^0(x)\over dx}
}
where $x = (\nu - \nu_0)/\DnuD$ is the dimensionless frequency shift from line
center.  The last approximation is valid for all $x$ when $\xb \ll 1$.  Because
of the incoherence of thermal emission, the Stokes parameters are simple
superpositions of the component intensities
\eqarray{
   I(x) &= &\fourth[(I^+ + I^-)(1 + \cos^2\theta) + 2I^0\sin^2\theta]
        \simeq I^0(x) \non
   V(x) &= &\half(I^+ - I^-)\cos\theta
        \simeq \xb {dI^0(x)\over dx}\cos\theta
}
where the final expressions are the terms to leading order in \xb.  Therefore,
\eq{\label{thermal}
        V(\nu) = \DnuB\cos\theta\,I'(\nu), \qquad \hbox{i.e. }\
                         \R(\nu) = \DnuB\cos\theta
}
(note that $\xb d/dx = \DnuB d/d\nu$). The constancy of the ratio profile \R\
reflects the symmetry of equation \ref{shift} --- the three $I^{\Delta m}$ are
described by a single spectral profile at appropriately shifted arguments.  This
symmetry is unique to thermal radiation.  The general symmetry of the problem is
invariance under mirror reflections perpendicular to the magnetic axis and its
consequence is instead
\eq{\label{mirror}
            I^+(\nu_0 + \delta\nu) =  I^-(\nu_0 - \delta\nu)
}
for any $\delta\nu$.  Equation \ref{shift} follows only if the spectral shape of
each $I^{\Delta m}$ is additionally symmetric about its centroid at $\nu_0 -
\Delta m\DnuB$.  This is the case for thermal radiation because the emission is
invariant under reversal of the particle motions.

\section{MASER CIRCULAR POLARIZATION}

In contrast with thermal emission, the Stokes parameters of maser radiation are
dominated by the interaction terms in eq.\ \ref{basic} rather than the source
terms when $J \ga \sqrt{S_0J_s}$ (\[E96]).  This condition is obeyed by
virtually all bright maser sources.  The polarization properties of radiation
generated in spontaneous decays and in stimulated emissions are entirely
different from each other. Consider for example a photon produced in a $\Delta m
= 0$ decay and thus linearly polarized. This photon contributes only to the
linear polarization of the source terms and will be reflected as such in the
polarization properties of observed thermal radiation.  However, this photon can
induce transitions with any value of $\Delta m$, thus any polarization, because
it can also be described as a coherent mixture of two circularly polarized
photons.  Consider, for instance, the stimulated emissions induced by this
photon in the case of a spin $1 \to 0$ transition. When the interacting particle
is in the $m = 0$ state, the induced photon is linearly polarized because it,
too, is generated in a $\Delta m = 0$ transition.  But when the particle is in
one of the $|m| = 1$ states, the induced photon is circularly polarized because
it is generated in a $\Delta m = \pm1$ transition (the interaction amplitude is
reduced, though).  A photon generated in stimulated emission has the same wave
vector as the parent photon but not necessarily the same polarization, because
the latter depends also on the magnetic quantum number of the interacting
particle.  Because of the finite line widths, different $\Delta m$ transitions
overlap when $\xb \ll 1$ and particles in the same magnetic sub-level can
interact with photons of different polarizations.  Amplification by stimulated
emission mixes the original polarizations of all the photons created in
spontaneous decays.

The polarization mixing effect of the amplification process is reflected in the
linear combinations in eqs.\ \ref{kappas} and \ref{R} and is the reason for the
rotation of mode polarization.  After amplification, any polarization the
original radiation might have had is irrelevant.  For example, even for thermal
emission in a magnetic field polarized according to eq.\ \ref{thermal}, each
mode contains random linear polarization (arbitrary $q$ and $u$) for propagation
in any direction other than $\theta = 0$.  Because the projection of every \PI\
on the $q$--$u$ plane points at a random direction, each mode rotates
differently during the amplification process, individual polarizations are
randomized, and the initial overall polarization disappears once the radiation
field is dominated by the induced photons.  The only polarization that can
survive the amplification process is that of stationary modes that do not
rotate, locking the individual polarization vectors.  All four Stokes parameters
then grow at the same rate and the fractional polarization remains constant. The
stationary modes of amplified radiation are the eigenvectors of the matrix \bfK\
and were identified in \[E96].  Denote by $\ko(x)$ the unsaturated absorption
coefficient of the maser transition in the absence of magnetic fields.  In the
presence of a magnetic field, the unsaturated absorption coefficient of the
$\Delta m$ transition becomes
\eq{\label{k's}
               \k_0^{\Delta m}(x) = \ko(x + \xb\Delta m)
               \simeq \ko(x) + \xb\Delta m{d\ko(x)\over dx},
}
similar to equation \ref{shift} for the corresponding intensities. Again, the
last relation is appropriate for $\xb \ll 1$.  Maser polarization properties are
controlled by the two ratios
\begin{eqnarray}
    \RI &= &{\k_0^+ + \k_0^- \over 2\koo}\ \simeq 1
                                                               \non
    \Rc &= &{\k_0^+ - \k_0^- \over 2\koo}\
         \simeq  {\xb\over\ko(x)}{d\ko(x)\over dx}.
\end{eqnarray}
In each case, the last expression is the result to leading order in \xb, with
the corrections smaller by order $x_B^2$ because of symmetry. The solution for
the maser circular polarization when $\xb \ll 1$ is
 \eq{\label{v}
      v_0(\nu) = \frac{8\Rc}{(2\RI + 1)\cos\theta}
      \simeq \frac{8\DnuB}{3\cos\theta} {\dko(\nu)\over \ko(\nu)},
 }
where $\theta$ is restricted by $\sqrt{2/3} \ge \cos\theta \ge 4\sqrt2\xb x$
(see \[E96], eqs.\ 4.13--4.15).  The last expression is obtained by inserting
\RI\ and \Rc\ from the previous equation, corrections are smaller by order
$x_B^2$.

\subsection{Unsaturated masers}

When the maser is unsaturated ($J(\nu) \ll J_s$), to leading order in both
$J/J_s$ and \xb, $\km \simeq \ko$ and \kl\ = \kc\ = 0 for any mode, as is
evident from equations \ref{kappas} and \ref{k's}.  In this approximation the
polarization rotation is neglected and the intensity follows the familiar
solution of an unsaturated scalar maser $I(\nu) = S_0 \exp\tau(\nu)$, where
$\tau(\nu) = \ko(\nu)\ell$ is the amplification optical depth, irrespective of
polarization.  Therefore $I'/I = \tau\dko/\ko$, and for any polarization $v$
\eq{\label{Rapprox}
    \R(\nu) = \frac{v(\nu)}{\tau(\nu)}{\ko(\nu)\over\dko(\nu)}.
}
Inserting the stationary circular polarization from equation \ref{v}, the
\R-profile of the maser solution in the unsaturated regime is
\eq{\label{unsaturated}
               \R(\nu) = \frac{8\DnuB}{3\cos\theta}\Psi(\nu),
}
where
\eq{\label{Psi}
 \Psi(\nu) =
{1\over\tau(\nu)} = {1\over \tau_0 + \displaystyle\ln{I(\nu)\over I(\nu_0)}}.
}
The profile $\Psi(\nu)$ expresses the spectral shape of (the unmeasurable)
$\tau(\nu)$ in terms of the measured intensity $I(\nu)$ and the free parameter
$\tau_0 = \tau(\nu_0)$, the maser optical depth at line center. This profile
increases toward the line wings from a  minimum at line center, fundamentally
different from the flat profile of thermal  radiation.  When such \R\ spectral
shape is detected the maser is unsaturated, therefore the spectral profile of
ln$I(\nu)$ can be used to deduce the profile of $\ko(\nu)$.  The \ko\ profile
determines the intrinsic Doppler linewidth, a quantity which is not known
a-priori and which is yet to be reliably determined in any maser source.
Furthermore, the spectral relation between $\R(\nu)$ and ln$I(\nu)$ provides a
method for direct determination, the first of its kind, of the maser optical
depth $\tau_0$ from observable quantities.  Once $\tau_0$ is found, ${\cal
R}(\nu_0)$ can be used to determine $B/\!\cos\theta$.

It may be noted that the polarization vector of the maser solution is not fully
stationary during the early stages of maser growth when $J/J_s < x_B^2$; indeed,
no polarization mode can avoid rotation at that stage.  An estimate of this
early rotation can be obtained by including the terms linear in \xb\ while
maintaining zeroth order in $J/J_s$. In this approximation order \km\ remains
equal to \ko\ so the intensity growth is the same, and \kl\ still is zero so the
linear polarization does not rotate. However, $\kc \ne 0$ and the circular
polarization does vary according to
\eq{\label{rotation}
       {dv \over d\tau} = \left(1 - v^2 \right)\Rc\cos\theta.
}
The solution of this equation for any initial polarization $v_i$ is
\eq{
     v = {v_i + \tanh\tau\Rc\cos\theta \over 1 + v_i\tanh\tau\Rc\cos\theta}\,,
}
a solution valid as long as $\exp(\tau_0)S_0/J_s < x_B^2$.  Pumping schemes of
OH masers typically have $S_0/J_s \sim 10^{-5}$ and the observed 20\%
polarization is reproduced with \xb\ = 0.03 for propagation at $\cos^2\theta =
1/3$. Therefore this solution holds for $\tau_0 \la \ln(100) = 4.6$. Figure 1
displays the \R-profiles of the polarization solution ($v_i = v_0$) at $\tau_0$
= 5 with and without the rotation.  The difference between the two is
practically insignificant.  Furthermore, note that the neglected terms of higher
order in $J/J_s$ will push the polarization toward stationary behavior and thus
decrease the variation of $v$ below the result of equation \ref{rotation}.
Therefore, this approximation produces an upper limit on variation of $v$ and
the actual solution must fall between the two displayed profiles. As $\tau_0$
increases, the linear polarization joins the rotation so that the overall
rotation of the polarization solution decreases further in inverse proportion to
$J$ and the stationary approximation becomes even better.

The deviation from flatness of the maser \R\ profile is a general result,
reflecting line narrowing during unsaturated growth (e.g., \[Book] \S4.5). It is
instructive to examine the unsaturated amplification of polarized thermal
radiation with an arbitrary intensity profile $\Ith(\nu)$ and a constant \R\
profile, i.e., $V \propto I'$. Equal amplification of the Stokes parameters
produces
\eq{
     I(\nu) = \Ith(\nu)e^{\tau(\nu)}, \qquad
     V(\nu) = \DnuB\cos\theta\,I'_{\rm th}(\nu)e^{\tau(\nu)}.
}
Evidently, $V$ can not remain proportional to $I'$ because the derivative of the
amplified intensity is not equal to the amplified intensity derivative. The
proportionality would be retained only if $\tau(\nu)$ were independent of $\nu$
instead of being sharply peaked at line center. Since the spectral shape of
\Ith\ follows the same Doppler profile as \ko, straightforward algebra yields
\eq{
              \R(\nu) = {\DnuB\cos\theta \over 1 +\tau_0\exp(-x^2)}\,.
}
The \R\ profile is constant only at $\tau_0 = 0$, i.e., only for the input
radiation. The amplification process destroys the profile flatness even though
it only amplifies the intensity without affecting the polarization. When $\tau_0
> 1$, \R\ assumes instead a Gaussian absorption shape.  This happens because the
amplification is centered on $\nu_0$, therefore the $I^{\pm}$ components are
amplified more strongly on their inner shoulders, the ones closer to $\nu_0$.
While this preserves the general symmetry principle of eq.\ \ref{mirror}, as it
must, the invariance under frequency shifts of eq.\ \ref{shift} is destroyed
because $I^{\pm}(\nu)$ are no longer symmetric about their centroids at $\npm =
\nu_0 \pm \DnuB$.

\subsection{Saturated masers}

When the maser is saturated ($J(\nu) > J_s$), the radiative transfer equation
for the intensity of the dominant ray of the polarization solution becomes
\eq{
          {dI(\nu)\over d\ell} = 3\pi J_s\frac{\ko(\nu)}{\Onu},
}
where \Onu\ is the maser beaming angle, which in general may vary with frequency
shift from line center (\[E90]). Therefore, the profile of $I(\nu)$ has the
spectral shape of the ratio $\ko(\nu)/\Onu$.

The frequency variation of \Onu\ depends on the type of maser amplification,
which is controlled by the geometry (\[EHM]). In {\em matter-bounded} masers,
whose prototypes are filaments, the beaming angle is independent of frequency
and $I(\nu) \propto \ko(\nu)$. In {\em amplification-bounded} masers the beaming
angle varies with frequency shift from line center according to dimensionality.
In the prototype planar maser, a saturated disk, $\Onu \propto 1/\ko(\nu)$ and
$I(\nu) \propto [\ko(\nu)]^2$. And in the prototype three-dimensional maser, a
saturated sphere, $\Onu \propto 1/[\ko(\nu)]^2$, so $I(\nu) \propto
[\ko(\nu)]^3$.  In summary, the intensity spectral profile of a saturated maser
obeys
\eq{\label{Isat}
                         I(\nu) \propto [\ko(\nu)]^p
}
where $p$ is the dimensionality of the geometry: 1 for filaments, 2 for planar
masers and 3 for sphere-like configurations.  Therefore $I'/I = p\dko/\ko$ and
the \R-profile of a saturated maser is
\eq{\label{saturated}
                         \R(\nu) = \frac{8\DnuB}{3p\cos\theta}.
}
The unsaturated profile $\Psi$ is replaced by the constant $1/p$ during the
saturation process, reflecting the rebroadening of the maser line.  Similar to
thermal radiation, \R\ is constant across the saturated maser line, only the
value of this constant is different. For a given magnetic field and propagation
at the smallest angle allowed for maser polarization, $\theta =
\cos^{-1}\sqrt{2/3} = 35^\circ$, the proportionality constant for filamentary
masers is 4 times larger than for thermal radiation (twice as large for disks,
1.3 times for spheres). In other words, masers require smaller fields to produce
the same circular polarization as thermal radiation.  The disparity between the
two cases increases with $\theta$ in proportion to $1/\cos^2\theta$.

The saturation effect was introduced here in its standard form.  Strictly, this
form applies only to linear masers, three-dimensional effects lead to a more
complex form (\[Lit73], \[BeK], \[Neu]).  However, because of the tight beaming
of maser radiation, the standard form provides an adequate approximation around
line center, failing only in the extreme wings ($x > 2$; \[E94]) of saturated
masers, and its use is justified in the present analysis.  In addition,
frequency redistribution has been neglected.  Incorporation of this important
ingredient into maser theory is sufficiently difficult that it has not yet been
fully accomplished even for scalar masers.  In keeping with common practice, it
has been neglected in this first study.  Also, the maser polarization problem
was solved only for a $J = 1 \to 0$ transition.  However, when $\xb \ll 1$ the
solution holds for all spins (E96) and thus is applicable here.  Note, though,
that this result holds only for isotropic pumping.  Pumping schemes that will
introduce $m$-dependence would require separate handling.

\section{RESOLVED ZEEMAN PATTERN; $\xb > 1$}

When $\xb > 1$ the Zeeman components separate and the radiation displays three
fully polarized lines whose linear and circular fractional polarizations are
constant across each component.  The circular polarizations of the two
$\sigma$-components, centered on $\npm = \nu_0 \pm \DnuB$, are
\eq{
           \vpm = \pm{2\cos\theta \over 1 + \cos^2\theta}
}
whatever the degree of saturation (\[GKK]; \[E96]; see \[Field] for the approach
to this solution of amplified unpolarized background radiation).  In fact, this
result follows from general symmetry properties and holds also for thermal
radiation, as can be easily verified from expressions listed in \[Cru].
Although the polarization is constant across each individual profile, it
maintains the reflection symmetry of eq.\ \ref{mirror} since $v^- = - v^+$.
Independent of the nature of the radiation, $V(\nu)$ is proportional to $I(\nu)$
across each component, not to $I'(\nu)$. The ratio profile now obeys $\R = \vpm
I/I'$, leading to
\eq{
  \R(\nu) = -{\vpm(\DnuD)^2 \over 2(\nu - \npm)}\times\cases
                  {1          & \hbox{thermal}              \cr
                   \Psi(\nu)  & \hbox{unsaturated maser}    \cr
                   1/p        & \hbox{saturated maser}      \cr}
}
if \ko\ and the thermal radiation follow a Doppler profile with width \DnuD.  In
the profile $\Psi$ of each unsaturated component, the intensity is normalized to
its line center magnitude for that component (cf.\ eq.\ \ref{Psi}).

The ratio profiles for $\xb \ll 1$ and $\xb > 1$ are fundamentally different
from each other.  In one case \R\ is symmetric in reflections about the center
of the single observed line, in the other it is anti-symmetric about the center
of each observed $\sigma$-component at \npm.  This difference provides a
decisive observational test to determine whether the Zeeman splitting is larger
or smaller than the linewidth.

\section{OBSERVATIONAL IMPLICATIONS}

Spectral analysis of circular polarization provides a new powerful tool for
maser studies.  Comparison of the profiles of $V(\nu)$ and $I'(\nu)$ offers
sensitivity to details at the level of \about\ \xb\ and has long been utilized
in studies of thermal radiation. The maser analysis can be readily performed
with the techniques developed for the thermal case (see e.g.\ \[Trol]) where the
observed $V$ spectrum is fitted as
\eq{
                         V(\nu) = aI(\nu) + bI'(\nu)
}
and $a$ and $b$ are free parameters.  The $aI$ term is introduced to account for
differences in instrumental sensitivity in left and right circular polarization.
The parameter $a$ has no bearing on the source properties, it is merely adjusted
to make $V - aI$ anti-symmetric about line center.  The next step is to fit $V -
aI$ to the intensity derivative $I'$, producing the meaningful parameter $b =
\DnuB\cos\theta$ (eq.\ \ref{thermal}).

In the case of maser radiation, various mechanisms can filter one sense of
circular polarization in the source itself (see e.g.\ \S 6.7 of \[Book] and
references therein). Then the observed $V$ profile may not be perfectly
anti-symmetric even before it reaches the detection instruments. Similar to
instrumental effects, the left-right asymmetry introduced by such filters can be
handled by fitting the observed $V$ profile of an unsaturated maser as
\eq{
       V(\nu) = aI(\nu) + {bI'(\nu) \over c + \ln [I(\nu)/I(\nu_0)]}
}
(eq.\ \ref{unsaturated}), with $a$ adjusted to make $V - aI$ anti-symmetric
about line center.  Fitting the resulting anti-symmetric profile produces the
adjustable parameters $b = 8\DnuB/3\cos\theta$ and $c = \tau_0$, the maser
optical depth at line center.  This is the first method to directly determine
the maser optical depth for self amplification and the intrinsic velocity
Doppler width (obtained from the spectral shape of ln$I$).

When the maser is saturated \R\ is flat, so the $V$ profile is fitted as in the
thermal case only $b = 8\DnuB/3p\cos\theta$ (eq.\ \ref{saturated}). The
parameter $p$ is determined by the maser geometry (it is 1 for filaments, 2 for
planar masers and 3 for three dimensional masers), and this cannot be found
directly from observations.  However, strong masers are unlikely to be
spherical, so $p$ = 3 can usually be dismissed.  In particular, the relevant
geometry for shock generated masers, applicable to water masers as well as the
1720 MHz masers at the Galactic center (\[YZ95]; \[Frail]; \[YZ96]), is likely
to be planar or filamentary and $p$ is either 2 or 1.  Another handle on $p$ can
come from the linewidth \Dnum\ of the observed maser intensity.  When \ko\ has a
Doppler shape with width \DnuD, equation \ref{Isat} shows that
\eq{
                         \Dnum = {\DnuD\over\sqrt{p}}
}
and the linewidth observed in disk masers is only 70\% of the Doppler width.  A
line narrower than the expected thermal width in a saturated maser can be taken
as indication of planar rather than filamentary geometry.

The distinct shape of the unsaturated \R\ profile, evident in figure 1, provides
the first direct method to determine the saturation stage of a maser.  Detection
of the predicted logarithmic variation across the line would provide a unique,
unambiguous signature of unsaturated maser operation.  However, a flat \R\
cannot immediately be taken as conclusive evidence for saturated operation;
within the observational errors, such behavior can always be attributed to
sufficiently large $\tau_0$. If \D\ denotes the dynamic range of intensity
measured across an unsaturated maser line, the expected relative variation of
\R\ is (1/$\tau_0$)ln\D.  The observed profile could be indistinguishable from a
constant if this variation is smaller than the observational error $\epsilon$,
namely, if
\eq{\label{lower}
                         \tau_0 > {1\over\epsilon} \ln\D.
}
A flat \R\ only implies a {\em lower} limit on $\tau_0$.  However, if the maser
is indeed unsaturated, then there is also an {\em upper} limit on $\tau_0$.  In
this case, the measured brightness temperature at line center is $T_b =
T_x\exp\tau_0$, where $T_x$ is the maser excitation temperature.  This quantity
in turn is related to the fractional inversion $\eta = (n_2 - n_1)/(n_2 + n_1)$
via $T_x = h\nu_0/2k\eta$ (e.g.\ \[Book] \S4.2). Therefore $\tau_0 = \ln(2\eta
kT_b/h\nu_0)$ and an absolute {\em upper} limit on $\tau_0$ is obtained by
inserting in this expression $\eta = 1$, the maximum theoretical value, yielding
\eq{\label{upper}
                         \tau_0 < \ln{2kT_b \over h\nu_0}.
}
The optical depth of an unsaturated maser is always bounded by the observed
brightness temperature.  The maser can display a flat \R\ profile only if this
upper limit is compatible with the previous lower limit.

To summarize the observation analysis procedure: The free parameter $a$ is
varied until the combination $V - aI$ of the observed $V$ and $I$ profiles is
anti-symmetric about line center, then the ratio $I'/(V - aI)$ is formed.  If
this ratio displays linear variation with ln$I$ across the line then the maser
is unsaturated and the slope and intercept of the linear fit determine
$B/\!\cos\theta$ and $\tau_0$.  If, instead, the ratio $I'/(V - aI)$ is constant
across the line then the situation depends on the last two bounds.  If these
bounds are incompatible, the maser is saturated.  That is, if the fitting error
$\epsilon$ and the observed brightness temperature $T_b$ and dynamic range \D\
obey
\eq{
                    \ln\D > \epsilon\ln{2kT_b \over h\nu_0}
}
yet $I'/(V - aI)$ is constant across the line then the maser must be saturated.
When the reverse relation exists between \D, $\epsilon$ and $T_b$, the
saturation stage is unsettled; the data lack the sensitivity to distinguish
between the \R\ profiles of saturated and unsaturated maser operation.

Observational data that obey the last inequality are of sufficient quality to
determine conclusively whether the maser is saturated or not. The recent VLA
polarization data of OH 1720 MHz masers near the Galactic center by \[YZ96] are
the first to meet this criterion.  The brightness temperatures of all resolved
Galactic OH masers ($h\nu_0/k$ =  0.08 K) always obey $T_b \ga 10^{10}$ K.  If
any such maser is unsaturated, the upper bound of eq.\ \ref{upper} implies that
$\tau_0 < 26$ (a realistic estimate for the inversion, $\eta \la 10\%$, would
reduce this bound to $\tau_0 < 24$).  The \R\ profiles of the Galactic-center
masers were fitted as constants by \[YZ96] and the errors listed for the
magnetic fields are less than 10\% in all sources, as low as 3\% in the
strongest (source A).  These are essentially error estimates of the quality of
the fits to flat ratio profiles, thus $\epsilon \sim$ 0.03--0.1. The resulting
fits match the observations all the way to the noise level of \about\ 10 mJy for
an intensity dynamic range $\D \ga 1000$ in the strongest source and no less
than \about\ 20 in the weakest (source F). Therefore, from equation \ref{lower}
the lower limits on $\tau_0$ range from 30 to 230 and are incompatible with the
upper limit for unsaturated OH masers.  This conclusion can be strengthened by a
proper analysis that would test the hypothesis of an unsaturated \R\ spectral
shape with the observations.  Such analysis should produce constraints that are
even tighter and more secure. Still, it seems safe to conclude already on the
basis of the current analysis that at least the strongest masers, and quite
possibly all of them, are saturated.  Previously, the strongest evidence for
saturation was provided by OH 1612 MHz masers in late-type stars (\[Book], \S
8.6).  This maser emission follows the temporal variation of the IR radiation,
the presumed pump, and the similar amplitudes indicate a linear relation between
pump and maser intensity, as expected in saturated operation.  While this
evidence seems persuasive, it is indirect as it requires an assumption, however
plausible, regarding the pump mechanism.  In contrast, the new evidence is
direct, obtained from the relation between $I$ and $V$, two maser intensities
measured independently.

Detailed profile analysis of circular polarization also enables direct,
unambiguous determination of the relation between the magnitudes of the Zeeman
splitting and the line Doppler width.  Substantial circular polarization (50\%
and higher) has long been observed in OH maser emission from late-type stars
(e.g., \[Reid], \[Clau], \[Cohen]).  In the absence of theory for $\xb < 1$ at
the time, this was taken as a signature of $\xb \ga 1$, implying magnetic fields
of at least \about\ 1--10 milligauss. However, this polarization often appears
as sharp reversals between adjacent narrow spectral components of the parameter
$V$, as expected for $\xb < 1$.  Analysis in terms of the theory for $\xb < 1$
yields instead fields of only \about\ 0.1 milligauss (\[E96]), so it is
important to determine conclusively what is the relevant domain of \xb.  The new
profile analysis provides a simple, reliable method to resolve this issue.  If
the $V$ profile can be fitted across each component as $V = aI$ without
residuals then $\xb > 1$.  If, on the other hand, the residuals are significant
and $V - aI$ is anti-symmetric around line center then $\xb < 1$.

Finally, circular polarization analysis, of either thermal or maser radiation,
can never fully determine the magnetic field because the propagation angle
$\theta$ is not known.  However, in the case of maser radiation $\theta$ can be
determined from linear polarization measurements; the polarization solution
shows that both circular and linear polarizations are generated when $\xb \ll
1$, and to leading order in \xb\ the linear polarization is a unique function of
$\theta$ (\[E96], eq.\ 4.14). Indeed, strong linear polarization has been
recently reported for the 1720 MHz masers by \[Cl97].  Although Faraday rotation
can reduce the linear polarization, even eliminate it on occasion, study of
linear polarization is an important task for observations.

\acknowledgments

Discussions with Tom Troland and Farhad Yusef-Zadeh, and the partial support of
NASA grants NAG 5-3010 and NAG 5-7031 are gratefully acknowledged.


\begin{figure}
\centering \leavevmode \epsfxsize = \hsize \epsfclipon \epsfbox{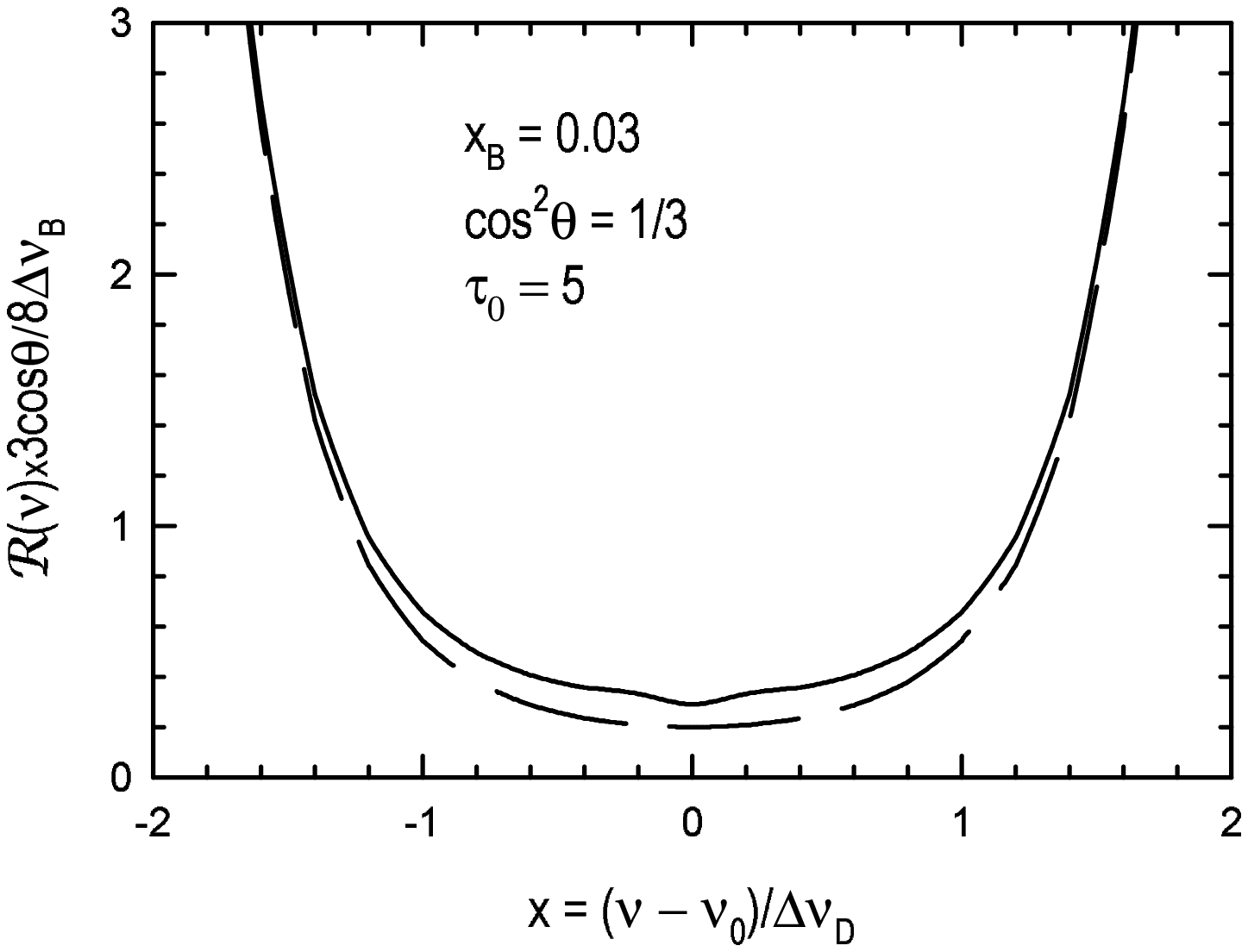}

\figcaption[fig1.ps]{The \R\ profile of the polarization solution for an
unsaturated maser with the parameters listed. The full line is the profile
including the maximum possible polarization rotation (eq.\ 21), the dashed line
is the profile neglecting this rotation (eq.\ 19).}

\end{figure}

\end{document}